\documentclass[12pt]{JHEP}

\title{Brane-World Cosmology in Higher Derivative Gravity or
Warped Compactification in the Next-to-leading Order of
AdS/CFT Correspondence}

\author{
Shin'ichi Nojiri \\
Department of Applied Physics \\
National Defence Academy, 
Hashirimizu Yokosuka 239, JAPAN \\
email: nojiri@cc.nda.ac.jp
}
\author{
Sergei D. Odintsov\footnote{On leave 
from Tomsk Pedagogical University, 634041 Tomsk, RUSSIA} \\
Instituto de Fisica de la Universidad de 
Guanajuato \\
Apdo.Postal E-143, 37150 Leon, Gto., MEXICO\\
email: odintsov@ifug5.ugto.mx, odintsov@mail.tomsknet.ru
}
\abstract{
The general model of higher derivative (HD) gravity is considered. 
The search of brane-world cosmology in such theory is presented when 
bulk is d5 AdS and boundary is spherical, hyperbolic or flat (single) 
brane. It is found the wide range of theory parameters where such
cosmology 
may be realized. Special attention is paid to the version of HD theory 
representing SG dual of ${\cal N}=2$ $Sp(N)$ SCFT (in next-to-leading 
order of large $N$ expansion). In particular, it is shown that 
inflationary brane Universe does not occur for SG dual while 
hyperbolic
brane occurs (which was not possible in leading order).

The quantum effects of CFT living on the brane (via the corresponding 
conformal anomaly induced effective action) may qualitatively change the
results of classical analysis. There appears inflationary (or hyperbolic) 
brane Universe induced by only quantum effects. In AdS/CFT correspondence 
(next-to-leading order) the addition of such CFT effective action (in some
energy region)
is naturally explained in terms of holographic renormalization group.
It results in the possibility of quantum creation of inflationary brane
Universe (with small rate) even for SG dual.
}

\begin{document}
\tolerance=5000
\def\pp{{\, \mid \hskip -1.5mm =}}
\def\cL{{\cal L}}
\def\be{\begin{equation}}
\def\ee{\end{equation}}
\def\bea{\begin{eqnarray}}
\def\eea{\end{eqnarray}}
\def\tr{{\rm tr}\, }
\def\nn{\nonumber \\}
\def\e{{\rm e}}
\def\D{{D \hskip -3mm /\,}}

\def\SEH{S_{\rm EH}}
\def\SGH{S_{\rm GH}}
\def\AdS5{{{\rm AdS}_5}}
\def\S4{{{\rm S}_4}}
\def\gfv{{g_{(5)}}}
\def\gfr{{g_{(4)}}}
\def\SC{{S_{\rm C}}}
\def\RH{{R_{\rm H}}}

\section{Introduction}

In brane-world scenarios one lives on the boundary 
(observable Universe) where gravity is trapped \cite{RS}.
Such brane is embedded in higher dimensional bulk space. The 
investigation of cosmological aspects of brane-worlds \cite{CH,cosm}
(see also references therein) shows that at some circumstances  
the inflationary Universe could be realized on the brane.
Even the experimental tests to search for higher dimensional 
deviations of our world due to bulk/boundary structure may 
be proposed 
\cite{exp}. However, the study of such brane cosmology has been done so
far almost exclusively for Einstein or dilatonic gravities. On the same 
time, higher derivative (HD) gravity represents very natural
generalization of general relativity. It enjoys various nice features 
like renormalizability \cite{ste} and asymptotic freedom in four
dimensions (see \cite{BOS} for introduction and review), possibility 
of self-consistent compactification on quantum and classical levels 
(see examples in \cite{BBO}), sufficiently small corrections to Newton
potential at reasonable range of parameters, etc. Moreover, HD terms 
are typical for string effective action in the derivatives expansion
\cite{GSW,Pol}. Hence, HD gravity in higher dimensions represents 
the interesting model where brane-world cosmology should be investigated 
and the trapping of gravity should be discussed. Note that propagator in
such theory is qualitatively different from the one in general relativity 
that is why some new phenomena may be expected.

The important remark is in order. In the widely accepted versions of 
brane-world scenario \cite{RS,CH,cosm} one studies $d+1$-dimensional gravity 
coupled to a brane in the formalism where normally two free parameters 
(bulk cosmological constant and brane tension) present. Adding of
$d$-dimensional Gibbons-Hawking boundary action and brane cosmological
constant term to action one can get de Sitter brane in AdS bulk even in
Einstein gravity. 
The position of such brane is fixed in terms of brane tension. 

Our ideology is somehow different, in the spirit of refs.\cite{HHR,NOm}.
Namely, one considers the addition of surface counterterms which make
variational procedure to be well-defined and eliminate the leading
divergences of action. Brane cosmological constant (or brane tension) 
is not considered as free parameter (as it was in original brane-world
scenario) but it is fixed by condition of finiteness of spacetime when
brane goes to infinity. In such approach, the possibility of cosmological
de Sitter brane-world in Einstein theory is eliminated. However, as we
explain below  in theories different from Einstein gravity this 
possibility is not ruled out thanks to other free parameters of theory.

Our purpose in the present work will be to study the brane-world cosmology 
in d5 HD gravity without (or with) quantum corrections. We consider
general model with arbitrary coefficients (next section) and derive bulk
and boundary equations of motion. The explicit structure of surface
counterterms is very important in such derivation. 
 
The solution of bulk equation of motion gives d5 AdS. Then, the brane
equation of motion
is discussed. It gives the restrictions to HD gravity parameters from
the condition of realization of spherical, hyperbolic or flat branes
on the boundary. The corresponding radius is derived when it exists. 
The specific versions of HD gravity like Weyl or Gauss-Bonnet theories
appear in this formalizm as particular examples.

It is very important that there may be twofold point of view to HD gravity.
 From one side, this is just alternative to general relativity. From
another side, within AdS/CFT correspondence \cite{AdS} some versions of HD
gravity represent SG duals where Einstein and cosmological terms are of
leading order and HD terms are of next-to-leading order in large $N$
approximation. The explicit example of that sort is presented in next
section (SG dual to ${\cal N}=2$ SCFT). Brane-world cosmology for such
theory
naturally appears as warped compactification in the next-to-leading order 
of AdS/CFT correspondence. It is shown that in such situation the creation
of spherical brane is impossible in leading as well as in next-to-leading
order of large $N$ expansion. On the same time, the account of
next-to-leading order terms makes possible the creation of hyperbolic 
brane living in d5 AdS bulk (this effect was prohibited in the leading
order of AdS/CFT correspondence). 

In section three we investigate the modification of above scenario 
when quantum CFT is living on the brane. This is done 
via the anomaly induced effective action. (The corresponding
study for
bulk Einstein gravity has been done in 
refs.\cite{sergio,HHR,NOm}.\footnote{In usual 4d world the anomaly driven inflation has been 
proposed in refs.\cite{sta}.}
In fact, this puts world-brane scenario in form of warped compactification
in AdS/CFT set-up  as the corresponding RG flow.) 
We show that for range of HD gravity parameters where classical
consideration does not give
inflationary or hyperbolic branes, the quantum brane matter effects
improve the situation: purely quantum creation of inflationary 
or hyperbolic branes in d5 AdS occurs (like in Einstein gravity).
On the same time if the phenomenon already existed on classical level
then quantum corrections do not destroy it. The bulk clearly is not
modified. 

For SG dual of ${\cal N}=2$ 
SCFT theory such picture may be naturally understood in terms of 
holographic RG description\cite{ver} in AdS/CFT set-up. That is two dual
descriptions (SG dual
and QFT dual) are matched together in some energy region into
the global representation of some RG flow. It is interesting that both
sides
are given in next-to-leading order of large $N$ expansion as 
conformal anomaly coefficients for SCFT under discussion include not
only quadratic but also linear terms on $N$. Then, it is shown that in 
such theory the spherical brane in d5 AdS is created (the role of
conformal
anomaly coefficients is dominant).  Without quantum CFT living on the
brane it was impossible. For hyperbolic brane creation the qualitative
results are not changed if compare with previous section.
Finally, in the last section we present brief summary of results 
and mention some possible developments along the direction under
consideration.

\section{Brane-World Cosmology in HD gravity}

We consider 5d spacetime whose boundary is 4 dimensional 
sphere S$_4$, which can be identified with a D3-brane, 
four-dimensional hyperboloid H$_4$, or four dimensional flat 
space R$_4$. 
The bulk part is given by 5 dimensional 
Euclidean Anti-de Sitter space $\AdS5$ 
\be
\label{AdS5i}
ds^2_\AdS5=dy^2 + \sinh^2 {y \over l}d\Omega^2_4\ .
\ee
Here $d\Omega^2_4$ is given by the metric of S$_4$, H$_4$ 
or R$_4$ with unit radius. One also assumes the boundary (brane) 
lies at $y=y_0$ and the bulk space is given by gluing two regions 
given by $0\leq y < y_0$.

One starts with the following action:
\be
\label{vi}
S=\int d^5 x \sqrt{\hat G}\left\{a \hat R^2 
+ b \hat R_{\mu\nu}\hat R^{\mu\nu}
+ c \hat R_{\mu\nu\xi\sigma}\hat R^{\mu\nu\xi\sigma}
+ {1 \over \kappa^2} \hat R - \Lambda \right\}\ .
\ee
Here the conventions of curvatures are 
given by
\bea
\label{curv}
R&=&g^{\mu\nu}R_{\mu\nu} \nn
R_{\mu\nu}&=& -\Gamma^\lambda_{\mu\lambda,\kappa}
+ \Gamma^\lambda_{\mu\kappa,\lambda}
- \Gamma^\eta_{\mu\lambda}\Gamma^\lambda_{\kappa\eta}
+ \Gamma^\eta_{\mu\kappa}\Gamma^\lambda_{\lambda\eta} \nn
\Gamma^\eta_{\mu\lambda}&=&{1 \over 2}g^{\eta\nu}\left(
g_{\mu\nu,\lambda} + g_{\lambda\nu,\mu} - g_{\mu\lambda,\nu} 
\right)\ .
\eea
When $a=b=c=0$, the action (\ref{vi}) becomes that of the 
Einstein gravity:
\be
\label{viE}
S=\int_{M_{d+1}} d^{d+1} x \sqrt{\hat G}\left\{
{1 \over \kappa^2} \hat R - \Lambda \right\}\ .
\ee
If we choose 
\be
\label{viWa}
a={1 \over 6}\hat c \ ,\quad b= -{4 \over 3}\hat c\ ,\quad
c=\hat c
\ee
the HD part of action is given by the square of the Weyl tensor 
$C_{\mu\nu\rho\sigma}$ : 
\be
\label{viWb}
S=\int d^5 x \sqrt{\hat G}\left\{
\hat c \hat C_{\mu\nu\xi\sigma}\hat C^{\mu\nu\xi\sigma}
+ {1 \over \kappa^2} \hat R - \Lambda \right\}\ .
\ee
It is interesting that 
the string theory dual to ${\cal N}=2$ superconformal field 
theory is presumably IIB string on ${\rm AdS}_5\times
X_5$ \cite{AFM} where $X_5=S^5/Z_2$. (The ${\cal N}=2$ $Sp(N)$ 
theory arises as the low-energy theory on the world volume on 
$N$ D3-branes sitting inside 8 D7-branes at an O7-brane). 
Then in the absence of Weyl term, 
${1 \over \kappa^2}$ and $\Lambda$ are given by 
\be
\label{prmtr}
{1 \over \kappa^2}={N^2 \over 4\pi^2}\ ,\quad
\Lambda= -{12N^2 \over 4\pi^2}\ .
\ee
This defines the bulk gravitational theory dual to super YM theory 
with two supersymmetries.
The Riemann curvature squared term in the above bulk action 
may be deduced from heterotic string via heterotic-type I duality 
\cite{Ts2}, which gives ${\cal O}(N)$ correction:
\be
\label{abc}
a=b=0\ ,\quad c= {6N \over 24\cdot 16\pi^2}\ .
\ee
Hence, HD gravity with above coefficients defines SG dual 
of super Yang-Mills theory (with two supersymmetries) in
next-to-leading order of AdS/CFT correspondence \cite{AdS}.

Using field redefinition ambiguity \cite{GWT} 
one can suppose that there exists the scheme where 
$R_{\mu\nu\alpha\beta}^2$ may be modified to 
$C_{\mu\nu\alpha\beta}^2$ in the same way as in ref.\cite{Marc}. 
Then, the action (\ref{viE}) is presumably the bulk action 
(in another scheme) dual 
to ${\cal N}=2$ SCFT.

Let us start from the bulk equations of motion.
First we investigate if the equations of motion for the general action 
(\ref{vi}) have a solution which describes anti de Sitter space, 
whose metric is given by
\be
\label{ai}
ds^2=\hat G^{(0)}_{\mu\nu}dx^\mu dx^\nu 
= {l^2 \over 4}\rho^{-2}d\rho d\rho + \sum_{i=1}^4
\rho^{-1} \eta_{ij}dx^i dx^j \ .
\ee
When we assume the metric in the form  (\ref{ai}), 
the scalar, Ricci and Riemann curvatures are given by
\be
\label{aii}
\hat R^{(0)}=-{20 \over l^2}\ ,\ \ 
\hat R^{(0)}_{\mu\nu}=-{4 \over l^2}G^{(0)}_{\mu\nu}\ ,\ \ 
\hat R^{(0)}_{\mu\nu\rho\sigma}=-{1 \over l^2}
\left(G^{(0)}_{\mu\rho}G^{(0)}_{\nu\sigma}
-G^{(0)}_{\mu\sigma}G^{(0)}_{\nu\rho}\right)\ ,
\ee
which tell that these curvatures are covariantly constant. 
Then in the equations of motion from the action (\ref{vi}), 
the terms containing the covariant derivatives of the curvatures 
vanish and the equations have the following form:
\bea
\label{aiii}
0&=&-{1 \over 2}G^{(0)}_{\zeta\xi}
\left\{a \hat R^{(0)2} 
+ b \hat R^{(0)}_{\mu\nu}\hat R^{(0)\mu\nu}
+ c \hat R^{(0)}_{\mu\nu\rho\sigma}\hat R^{(0)\mu\nu\rho\sigma}
+ {1 \over \kappa^2} \hat R^{(0)} - \Lambda \right\} \nn
&& + 2a R^{(0)} R^{(0)}_{\zeta\xi} 
+ 2b \hat R^{(0)}_{\mu\zeta}{\hat R^{(0)\mu}}_\xi
+ 2c \hat R^{(0)}_{\zeta\mu\nu\rho}\hat R_\xi^{(0)\mu\nu\rho}
+ {1 \over \kappa^2} \hat R^{(0)}_{\zeta\xi}\ .
\eea
Then  substituting Eqs.(\ref{aii}) into (\ref{aiii}), one gets 
\bea
\label{aiv}
0&=&{80a \over l^4} + {16b \over l^4} \nn
&& + {8c \over l^4} - {12 \over \kappa^2 l^2}
-\Lambda\ .
\eea
The equation (\ref{aiv}) can be solved with respect to $l^2$ if 
\be
\label{aiva}
{144 \over \kappa^4}-16\left\{20a + 4b + 2c\right\}
\Lambda\geq 0
\ee
which can been found from the determinant in (\ref{aiv}).
Then we obtain\cite{NOch}
\be
\label{ll}
l^2=-{{12 \over \kappa^2}\pm \sqrt{
{144 \over \kappa^4}-16\left\{20a + 4b + 2c\right\}
\Lambda} \over 2\Lambda}\ .
\ee
The sign in front of the root in the above equation 
may be chosen to be positive which 
corresponds to the Einstein gravity ($a=b=c=0$).
For SG dual of ${\cal N}=2$ $Sp(N)$ theory, we find from (\ref{abc}) 
\be
\label{N2ll}
{1 \over l^2}=1 + {1 \over 24N} + {\cal O}\left(N^{-2}\right)\ .
\ee

Now, let us discuss the surface terms in HD gravity 
on the chosen background
\be
\label{i}
ds^2\equiv\hat G_{\mu\nu}dx^\mu dx^\nu 
= {l^2 \over 4}\rho^{-2}d\rho d\rho + \sum_{i=1}^d
\hat g_{ij}dx^i dx^j \ , \quad 
\hat g_{ij}=\rho^{-1}g_{ij} \ .
\ee
If the boundary of AdS$_5$ lies at 
$\rho=\rho_0$, the variation $\delta S$ 
contains the derivative of $\delta \hat g^{ij}$ with respect 
to $\rho$, which makes the variational principle ill-defined. 
In order that the variational principle is well-defined on the 
boundary, the variation of the action should be written in the 
form of
\be
\label{bdry1}
\delta S=\int d^5x \sqrt{\hat G}
\delta\hat g^{ij}\times \left(\mbox{eq. of motion}\right)
+ \int_{\rho=\rho_0} d^4 x \sqrt{\hat g}\delta\hat g^{ij}
\left\{\cdots\right\}
\ee
after using the partial integration. If we put 
$\left\{\cdots\right\}=0$ for $\left\{\cdots\right\}$ in 
(\ref{bdry1}), we could obtain the boundary condition. If 
the variation of the action on the boundary contains 
$(\delta \hat g^{ij})'$, however, we cannot partially integrate 
it with respect to $\rho$ on the boundary
to rewrite the variation in the form of (\ref{bdry1}) since 
$\rho$ is the coordinate expressing the direction perpendicular 
to the boundary. Therefore the ``minimum'' of the action is 
ambiguous. Such a problem was well studied by Gibbons and Hawking 
in \cite{3} for the Einstein gravity ($a=b=c=0$). The boundary 
term was added to the action, which cancels the variation : 
\be
\label{bdry2}
S_b^{\rm GH} = -{2 \over \tilde \kappa^2}
\int_{\rho=\rho_0} d^4 x \sqrt{\hat g}
 D_\mu n^\mu \ .
\ee
Here $n^\mu$ is 
the unit vector normal to the boundary.  
In the coordinate choice  (\ref{i}), the action (\ref{bdry2}) 
has the form 
\be
\label{bdry3}
S_b^{\rm GH} = -{2 \over \tilde \kappa^2}
\int_{\rho=\rho_0} d^4 x \sqrt{\hat g}
{\rho \over l}\hat g_{ij}\left(\hat g_{ij}\right)'\ .
\ee
Then the variation over the metric $\hat g_{ij}$ gives 
\be
\label{bdry4}
\delta S_b^{\rm GH} = -{2 \over \tilde \kappa}
\int_{\rho=\rho_0} d^4 x \sqrt{\hat g}
{\rho \over l}\left[ \delta \hat g^{ij}
\left\{-\hat g_{ik}\hat g_{il}\left(\hat g_{kl}\right)'
-{1 \over 2}\hat g_{ij}\hat g_{kl}\left(\hat g_{kl}\right)'
\right\}+ \hat g_{ij}\left(\delta\hat g_{ij}\right)'\right]\ .
\ee
 From the other side, the surface terms in the variation of 
the bulk Einstein action 
($a=b=c=0$ in (\ref{viE})) have the form 
\bea
\label{bdry5}
\delta S^{\rm Einstein}
&=&\int d^5x \sqrt{\hat G}
\delta\hat g^{ij}\times \left(\mbox{Einstein equation}\right)\nn
&& + {1 \over \kappa^2}
\int_{\rho=\rho_0} d^4 x \sqrt{\hat g}{2\rho \over l}
\left[ \hat g_{ij}' \delta \hat g^{ij} 
+ \hat g_{ij}\left(\delta \hat g^{ij}\right)'  \right]\ .
\eea
Then we find the terms containing 
$\left(\delta \hat g^{ij}\right)' $ in (\ref{bdry4}) and 
(\ref{bdry5}) are cancelled with each other. 

We also need the counterterms, besides Gibbons-Hawking 
term (\ref{bdry2}), to cancell the divergence coming from 
the infinite volume of AdS. Such a kind of counterterms can be 
given by the local quantities on the 4 dimensional boundary. 
In \cite{NOch}, the surface counterterms are discussed for 
higher derivative gravities in all detail. Note that they 
are relevant also for quantum cosmology\cite{HL}. 

We also add the surface terms $S_b^{(1)}$ corresponding to 
Gibbons-Hawking term (\ref{bdry2}) and $S_b^{(2)}$ 
which is the leading 
counterterm corresponding to the vacuum energy on the brane:
\bea
\label{Iiv}
S_b&=&S_b^{(1)} + S_b^{(2)} \nn
S_b^{(1)} &=& \int d^4 x \sqrt{\hat g}\left[
4\tilde a\hat R D_\mu n^\mu 
+ 2\tilde b\left(n_\mu n_\nu \hat R^{\mu\nu} 
D_\sigma n^\sigma + \hat R_{\mu\nu}D^\mu n^\nu \right) \right. \nn
&& \left. 
+ 8\tilde c n_\mu n_\nu \hat R^{\mu\tau\nu\sigma} D_\tau n_\sigma 
- {2 \over \tilde \kappa^2}D_\mu n^\mu \right] \nn
S_b^{(2)} &=& - \eta\int d^4 x \sqrt{\hat g} \ .
\eea
In \cite{NOch}, in order to cancell the leading order 
divergence, which appears when the brane goes to infinity, 
we got
\be
\label{eta}
\eta=-{32 T \over l^3} + {8 \over l\kappa^2} 
+ {4\tilde T \over l^3} - {2 \over l\tilde\kappa^2} \ .
\ee
Here 
\be
\label{TU}
T=10a + 2b + c\ ,\quad \tilde T=10\tilde a 
+ 2\tilde b + \tilde c\ .
\ee
Note that unlike to standard brane-world scenarios $\eta$ is not free
parameter.

The metric of $\S4$ with the unit radius is given by
\be
\label{S4metric1}
d\Omega^2_4= d \chi^2 + \sin^2 \chi d\Omega^2_3\ .
\ee
Here $d\Omega^2_3$ is described by the metric of 3 dimensional 
unit sphere. If we change the coordinate $\chi$ to 
$\sigma$ by 
\be
\label{S4chng}
\sin\chi = \pm {1 \over \cosh \sigma} \ , 
\ee
one obtains
\be
\label{S4metric2}
d\Omega^2_4= {1 \over \cosh^2 \sigma}\left(d \sigma^2 
+ d\Omega^2_3\right)\ .
\ee
On the other hand, the metric of the 4 dimensional flat 
Euclidean space is given by
\be
\label{E4metric}
ds_{\rm 4E}^2= d\zeta^2 + \zeta^2 d\Omega^2_3\ .
\ee
Then by changing the coordinate as 
\be
\label{E4chng}
\zeta=\e^\sigma\ , 
\ee
one gets
\be
\label{E4metric2}
ds_{\rm 4E}^2= \e^{2\sigma}\left(d\sigma^2 + d\Omega^2_3\right)\ .
\ee
For the 4 dimensional hyperboloid with the unit radius, 
the metric is given by
\be
\label{H4metric1}
ds_{\rm H4}^2= d \chi^2 + \sinh^2 \chi d\Omega^2_3\ .
\ee
Changing the coordinate $\chi$ to $\sigma$  
\be
\label{H4chng}
\sinh\chi = {1 \over \sinh \sigma} \ , 
\ee
one finds
\be
\label{H4metric2}
ds_{\rm H4}^2 = {1 \over \sinh^2 \sigma}\left(d \sigma^2 
+ d\Omega^2_3\right)\ .
\ee

Motivated by (\ref{S4metric2}), 
(\ref{E4metric2}) and (\ref{H4metric2}), one takes  
the metric of 5 dimensional space time as follows:
\be
\label{metric1}
ds^2=dz^2 + \e^{2A(z,\sigma)}\sum_{i,j=1}^4
\tilde g_{ij}dx^i dx^j\ ,
\quad \tilde g_{\mu\nu}dx^\mu dx^\nu\equiv l^2\left(d \sigma^2 
+ d\Omega^2_3\right)\ .
\ee
Here the coordinate $z$ is related the coordinate $\rho$ in 
(\ref{ai}) by 
\be
\label{zrho}
\rho=\e^{-{2z \over l}}\ .
\ee
Under the choice of metric in (\ref{metric1}), the curvatures 
have the following forms:
\bea
\label{crvtrs}
R_{zi zj}&=&\e^{2A}\left(-A_{,zz} - 
\left(A_{,z}\right)^2\right)\tilde g_{ij} \nn
R_{zAz\sigma}&=&-l^2\e^{2A}A_{,z\sigma}g^s_{AB} \nn
R_{\sigma A\sigma B}&=&\left(-l^2 \e^{2A}A_{,\sigma\sigma} 
-l^2 \e^{4A}\left(A_{,z}\right)^2\right)g^s_{AB} \nn
R_{ABCD}&=&\left(l^2 \e^{2A} -l^2 \e^{2A}
\left(A_{,\sigma}\right)^2
- l^4\e^{4A}\left(A_{,z}\right)^2\right)
\left(g^s_{AC}g^s_{BD} - g^s_{AD}g^s_{BC}\right) \nn
R_{zz}&=&4\left(-A_{,zz} - \left(A_{,z}\right)^2\right) \nn
R_{z\sigma}&=&-3A_{,z\sigma} \nn
R_{\sigma\sigma}&=&l^2l^2\e^{2A}\left(-A_{,zz} 
- 4\left(A_{,z}\right)^2\right) - 3A_{,\sigma\sigma} \nn
R_{AB}&=&\left( l^2\e^{2A}\left(-A_{,zz} 
- 4\left(A_{,z}\right)^2\right) - A_{,\sigma\sigma} 
- 2\left(A_{,\sigma}\right)^2 +2\right) g^s_{AB} \nn
R&=&-8A_{,zz} - 20\left(A_{,z}\right)^2 +l^{-2}\e^{-2A}
\left(-6A_{,\sigma\sigma} - 6\left(A_{,sigma}\right)^2 +6 
\right) \ .
\eea
Here $\cdot_{,\mu\nu\cdots }\equiv 
{\partial \over \partial x^\mu}{\partial \over \partial x^\nu}
\cdots (\cdot)$. 
Other curvatures except those obtained by permutating the 
indeces of the above curvatures vanish. We also write the metric 
of S$_3$ in the following form:
\be
\label{S3m}
d\Omega^2_3=\sum_{A,B=1}^3g^s_{AB}dx^A dx^B\ .
\ee
One gets that $n^\mu$ and the covariant derivative of $n^\mu$ 
are
\be
\label{Dn}
n^\mu=\delta_{\rho}^\mu\ ,\quad 
D_i n^j = \delta_i^j A_{,z}\ (\mbox{others}=0)\ .
\ee
Then the actions $S$ in (\ref{vi}) and $S_b$ in (\ref{Iiv}) 
have the following forms:
\bea
\label{SA}
S&=& l^4 \int d^5x \e^{4A}\sqrt{g^s}\left[
a\left\{64 \left(A_{,zz}\right)^2 
+ 320 A_{,zz}\left(A_{,z}\right)^2 + 400\left(A_{,z}\right)^4
\right.\right. \nn
&& + 36 l^{-4}\e^{-4A}\left(A_{,\sigma\sigma}\right)^2 
+ 72 l^{-4}\e^{-4A}A_{,\sigma\sigma}\left(A_{,\sigma}\right)^2 
+ 36 l^{-4}\e^{-4A}\left(A_{,\sigma}\right)^4 \nn
&& + l^{-2}\e^{-2A}\left(96 A_{,zz} +240\left(A_{,z}\right)^2\right)
\left(A_{,\sigma\sigma} + \left(A_{,\sigma}\right)^2 - 1
\right) \nn
&& \left.+ l^{-4}\e^{-4A}\left(-72A_{,\sigma\sigma} 
-72 \left(A_{,\sigma}\right)^2 +36 \right)\right\} \nn
&& + b\left\{20 \left(A_{,zz}\right)^2 
+ 64 A_{,zz}\left(A_{,z}\right)^2 + 80 \left(A_{,z}\right)^4
+ 18l^{-2}\e^{-2A}\left(A_{,z\sigma}\right)^2 \right. \nn
&& + 12 l^{-4}\e^{-4A}\left(A_{,\sigma\sigma}\right)^2 
+ 12 l^{-4}\e^{-4A}A_{,\sigma\sigma}\left(A_{,\sigma}\right)^2 
+ 12 l^{-4}\e^{-4A}\left(A_{,\sigma}\right)^4 \nn
&& + l^{-2}\e^{-2A}\left(12 A_{,zz} + 48 \left(A_{,z}\right)^2\right)
\left(A_{,\sigma\sigma} + \left(A_{,\sigma}\right)^2 - 1
\right) \nn
&& \left.+ l^{-4}\e^{-4A}\left(-12A_{,\sigma\sigma} 
-24 \left(A_{,\sigma}\right)^2 +12 \right)\right\} \nn
&& + c\left\{16 \left(A_{,zz}\right)^2 
+ 32 A_{,zz}\left(A_{,z}\right)^2 + 40 \left(A_{,z}\right)^4
+ 24 l^{-2}\e^{-2A}\left(A_{,z\sigma}\right)^2
\right. \nn
&& + 12 l^{-4}\e^{-4A}\left(A_{,\sigma\sigma}\right)^2 
+ 12 l^{-4}\e^{-4A}\left(A_{,\sigma}\right)^4 \nn
&& \left. + 24 l^{-2}\e^{-2A}\ \left(A_{,z}\right)^2
\left(A_{,\sigma\sigma} + \left(A_{,\sigma}\right)^2 - 1
\right)  + 12 l^{-4}\e^{-4A}\left(
\left(-2A_{,\sigma}\right)^2 + 1 \right)\right\} \nn
&& +{1 \over \kappa^2}\left\{\left( -8 
A_{,zz} - 20 \left(A_{,z}\right)^2\right) \right. \nn
&& \left.\left. +\left(-6 A_{,\sigma\sigma} 
- 6 \left(A_{,\sigma}\right)^2 
+ 6 \right)\e^{-2A}\right\} + \Lambda \right] \\
\label{SbA}
S_b &=& l^4 \int d^4x \e^{4A}\sqrt{g^s}\left[
16\tilde a \left\{\left( -8 
A_{,zz} - 20 \left(A_{,z}\right)^2\right) \right.\right. \nn
&& \left. +\left(-6 A_{,\sigma\sigma} 
- 6 \left(A_{,\sigma}\right)^2 
+ 6 \right)\e^{-2A}\right\} A_{,z} \nn
&& + 2\tilde b \left\{\left( -20 
A_{,zz} - 32 \left(A_{,z}\right)^2\right) 
+\left(-6 A_{,\sigma\sigma} 
 - 6 \left(A_{,\sigma}\right)^2 
+ 6 \right)\e^{-2A}\right\} A_{,z} \nn
&& \left. + 32\tilde c \left( - A_{,zz} - \left(A_{,z}\right)^2\right) 
+ \eta\right]
\eea
>From the variation over $A$, one obtains the following equation 
on the brane, which lies at $z=z_0$:
\bea
\label{breq}
\lefteqn{\delta\left(S + 2S_b\right)} \nn
&=& 2V_3l^4 \int d\sigma \e^{4A}\left[\left(-32 \tilde T 
+ 24 \tilde U\right) A_{,z}\delta A_{,zz} \right. \nn
&& + \left\{\left(32 T - 24 U-32 \tilde T 
+ 24 \tilde U\right) A_{,zz}
+ \left\{\left(32 T - 24 U - 96 \tilde T 
+ 72 \tilde U\right) \left(A_{,z}\right)^2 \right.\right. \nn
&& \left. +12 (U-\tilde U)l^{-2}\e^{-2A}\left( A_{,\sigma\sigma} 
+ \left(A_{,\sigma}\right)^2 - 1\right)
- {8 \over \kappa^2} + {8 \over \tilde \kappa^2}\right\}
\delta A_{,z} \nn
&& + \left\{\left(64T -128\tilde T + 96 \tilde U
\right)A_{,zz}A_{,z} \right. \nn
&& + \left(160 T - 128 \tilde T\right) \left(A_{,z}\right)^3 \nn
&& \left(48 T - 24 \tilde U\right)l^{-2}\e^{-2A}
\left( A_{,\sigma\sigma}  + \left(A_{,\sigma}\right)^2 - 1
\right) A_{,z} \nn
&& + \left( -36b -96c -12\tilde U\right)l^{-2}\e^{-2A}
A_{,z\sigma\sigma} + \left(-72b - 192c\right)l^{-2}\e^{-2A}A_{,\sigma}
A_{,z\sigma} \nn
&& \left.\left. + \left(-{8 \over \kappa^2} + {32 \over \tilde\kappa^2}
\right)A_{,z} - 4\eta \right\}\delta A \right]\ .
\eea
The factors $2$ in front of $S_b$ and $V_3$ come from the fact 
that we are considering two bulk space (corresponding to 
$B_5^{(1,2)}$ in \cite{HHR}) which have one common boundary 
(S$_4$ in \cite{HHR}) or brane.

Here $T$ and $\tilde T$ are defined in (\ref{TU}) and 
\be
\label{UU}
U=8a + b \ ,\quad \tilde U=8\tilde a + \tilde b
\ee
and $V_3$ is the volume of the unit 3 sphere:
\be
\label{V3}
V_3=\int d^3x_A\sqrt{g^s}=2\pi^2\ .
\ee
In order that the variational principle is well-defined, 
the coefficients of $\delta A_{,zz}$ and $\delta A_{,z}$ 
should vanish. For general $A$, only one solution is given by 
the Weyl gravity in (\ref{viWa}) 
\bea
\label{viWaB}
&& a={1 \over 6}\hat c \ ,\quad b= -{4 \over 3}\hat c\ ,\quad
c=\hat c \nn
&& \tilde a={1 \over 6}\tilde{\hat c} \ ,\quad 
\tilde b= -{4 \over 3}\tilde{\hat c}\ ,\quad 
\tilde c=\tilde{\hat c}\nn
&& \tilde \kappa^2 = \kappa^2\ .
\eea
This remarkable property of Weyl gravity indicates to some natural 
connection between such version of HD gravity and brane physics.

When $A$ is given by the AdS$_5$ and the brane is S$_4$, 
\be
\label{AS4}
A=\ln\sinh{z \over l} - \ln\cosh\sigma\ ,
\ee
Eq.(\ref{breq}) has the following form:
\bea
\label{breq2}
\lefteqn{\delta\left(S + 2S_b\right)} \nn
&=& 2V_3l^4 \int d\sigma {\sinh^4 {z_0 \over l}
\over \cosh^4\sigma}\left[\left(-32 \tilde T 
+ 24 \tilde U\right){\coth{z_0 \over l} \over l} 
\delta A_{,zz} \right. \nn
&& + \left\{-{64\tilde T \over l^2}\coth^2{z_0 \over l} 
+ {32(T - \tilde T) \over l^2} + {8 \over \tilde\kappa^2}
- {8 \over \kappa^2}\right\}\delta A_{,z} \nn
&& \left.+ \left\{ -{48\tilde U \over l^2\sinh^2{z_0 \over l}}
\left(-{8 \over \kappa^2} + {32 \over \tilde\kappa^2}
\right){\coth{z_0 \over l} \over l} - 4\eta \right\}\delta A 
\right]\ .
\eea
Then in order that the variational principle is well-defined, 
we obtain
\bea
\label{wldeqs}
0&=&-32 \tilde T 
+ 24 \tilde U \nn
0&=& \tilde T \nn
0&=& {32(T - \tilde T) \over l^2} + {8 \over \tilde\kappa^2}
- {8 \over \kappa^2}
\eea
or
\be
\label{wldeqs2}
0=\tilde T =\tilde U\ ,\quad 
{1 \over \tilde\kappa^2}={1 \over \kappa^2} - {4T \over l^2}\ .
\ee
The above results are consistent with those in \cite{NOch}. 
Then since $\eta$ in (\ref{eta}) is given by
\be
\label{eta2}
\eta={6 \over l\kappa^2} - {24T \over l^3}
\ee
the equation of motion (in terms of the coefficients) has 
the following form:
\be
\label{breqc}
0=\left({24 \over \kappa^2} + {32 T\over l^2}
\right){\coth{z_0 \over l} \over l} - {24 \over l\kappa^2} 
+ {96 T\over l^3}\ . 
\ee
For the pure Einstein case ($a=b=c=0$ or $T=0$), the equation 
(\ref{breqc}) reproduces the previous equation in \cite{HHR,NOm} 
by putting $\kappa^2 = 16\pi G$. 
In the pure Einstein case, there is no solution of Eq.(\ref{breqc}).
 Then
for the 
case, we need to add the quantum correction coming from 
the trace anomaly of the matter fields on the brane in 
order that the equation corresponding to (\ref{breqc}) has a 
non-trivial solution.

 In case of the higher derivative gravity 
in (\ref{breqc}), there can be a solution in general. 
The r.h.s. in Eq.(\ref{breqc}) goes to positive infinity when 
$z_0\rightarrow +0$ if ${24 \over \kappa^2} + {32 T\over l^2}>0$. 
On the other hand, the r.h.s. becomes ${128T \over l^2}$ when 
$z_0$ goes to positive infinity. Then if $T<0$, there can be a 
solution in (\ref{breqc}) without the quantum correction on the 
brane. As the r.h.s. is the monotonically increasing function 
of $z_0$, there is only one solution if $T<0$.
We should also note that there does not appear corrections 
from $R^2$ gravity terms for the Weyl gravity (\ref{viWa}), 
where $T=0$. 
For SG dual of ${\cal N}=2$ $Sp(N)$ 
theory, from (\ref{abc}), we find 
\be
\label{N2T}
T={6N \over 24\cdot 16\pi^2}\ .
\ee
As $T>0$, there is no classical solution for spherical brane.

If we rewrite (\ref{breqc}) as 
\be
\label{breqcC}
{\coth{z_0 \over l} \over l} ={ {24 \over l\kappa^2}
 - {96 T\over l^3} \over {24 \over \kappa^2} + {32 T\over l^2}}\ , 
\ee
the r.h.s. is the monotonically increasing function 
of the absolute value $|T|$ of $T$ if $T<0$. Since 
$\coth{z_0 \over l}$ is the monotonically decreasing 
function of $z_0$, the radius 
${\cal R}$ of S$_4$, which is given by
\be
\label{tldR}
{\cal R} = l\e^{\tilde A(y_0)} = l \sinh{z_0 \over l}\ ,
\ee 
decreases if $|T|$ 
increases when $T<0$ and $l$ is fixed. We should note that 
$l$ can be a function of $T$ since $l$ is given by (\ref{ll}), 
which is given in terms of $T$ as follows:
\be
\label{llT}
l^2=-{{12 \over \kappa^2}\pm \sqrt{
{144 \over \kappa^4}-32T\Lambda} \over 2\Lambda}\ .
\ee
If we fix $\Lambda$ instead of $l$, the situation becomes 
very complicated. 

Using ${\cal R}$ in (\ref{tldR}), we can rewrite 
Eq.(\ref{breqcC}) in the following form:
\be
\label{breqCC}
0=\left({24 \over l\kappa^2} + {32 T\over l^3}
\right)\sqrt{1+ {l^2 \over {\cal R}^2}} - {24 \over l\kappa^2} 
+ {96 T\over l^3} \ . 
\ee
For SG dual of ${\cal N}=2$ $Sp(N)$ theory, using (\ref{N2ll}) and 
(\ref{N2T}), one gets
\be
\label{breqCCC}
0=\sqrt{1+ {1 \over {\cal R}^2}} - 1
+ {1 \over 48 N}\sqrt{1+ {1 \over {\cal R}^2}}\left(5 
- {1 \over {\cal R}^2 +1}\right) + {22 \over 48N}  
+ {\cal O}\left(N^{-2}\right) \ . 
\ee
In this case, there is no any solution for ${\cal R}$. 
It is remarkable that warped compactification to spherical brane
is not realistic in leading (Einstein theory) as well as in
next-to-leading order of AdS/CFT correspondence.

Instead of the brane of S$_4$ in (\ref{AS4}), we can consider 
the brane of H$_4$, where $A$ is given by
\be
\label{AH4}
A=\ln\cosh{z \over l} - \ln\sinh\sigma\ .
\ee
By the similar calculation as for S$_4$, we again obtain the 
Eqs.(\ref{wldeqs2}) and (\ref{eta2}). The equation 
corresponding to (\ref{breqc}) has the following form:
\be
\label{breqcH}
0=\left({24 \over \kappa^2} + {32 T\over l^2}
\right){\tanh{z_0 \over l} \over l} - {24 \over l\kappa^2} 
+ {96 T\over l^3}\ . 
\ee
In case of pure Einstein gravity, there is no solution. 
When $z_0=0$, the r.h.s. in Eq.(\ref{breqcH}) becomes 
$- {24 \over l\kappa^2} + {96 T\over l^3}$, which can be 
regarded as negative. On the other hand, when $z_0$ goes to 
positive infinity, the r.h.s. becomes ${128T \over l^2}$.   
Then if $T>0$, which is different from the case of the S$_4$ 
brane, there can be a solution in (\ref{breqcH}) 
without the quantum correction on the brane. 
Rewriting Eq.(\ref{breqcH}) in the form 
\be
\label{breqcHH}
{\tanh{z_0 \over l} \over l} ={ {24 \over l\kappa^2} 
- {96 T\over l^3} \over {24 \over \kappa^2} + {32 T\over l^2}}\ ,
\ee
we find the radius 
${\cal R}_{\rm H}$ of H$_4$, which is defined by 
\be
\label{tldRH}
{\cal R}_{\rm H} = l\e^{\tilde A(y_0)} = l \cosh{z_0 \over l}\ ,
\ee
The radius ${\cal R}_{\rm H}$ is monotonically decreasing function 
of $|T|$ again if $T>0$ and $l$ is fixed since the l.h.s. in 
(\ref{breqcHH}) is the monotonically increasing function of $z_0$ 
and the r.h.s. is the monotonically decreasing function of $|T|$ 
if $T>0$. 

Using ${\cal R}_{\rm H}$ in (\ref{tldRH}), one can present 
Eq.(\ref{breqcHH}) in the following form:
\be
\label{breqCCHH}
0=\left({24 \over l\kappa^2} + {32 T\over l^3}
\right)\sqrt{1 - {l^2 \over {\cal R}_{\rm H}^2}} 
- {24 \over l\kappa^2} 
+ {96 T\over l^3}
\ . 
\ee
For SG dual of ${\cal N}=2$ $Sp(N)$ theory,  using (\ref{N2ll}) and 
(\ref{N2T}), we have
\be
\label{breqCCCH}
0=\sqrt{1 - {1 \over {{\cal R}_{\rm H}}^2}} - 1
+ {1 \over 48 N}\sqrt{1 - {1 \over {\cal R}^2}}\left(5 
- {1 \over {{\cal R}_{\rm H}}^2 - 1}\right) + {22 \over 48N}  
+ {\cal O}\left(N^{-2}\right) \ . 
\ee
For large ${\cal R}_{\rm H}$, Eq.(\ref{breqCCC}) has the 
following form:
\be
\label{breqCCCHH}
0=-{1 \over 2{{\cal R}_{\rm H}}^2} + {2 \over 3N} 
+ {\cal O}\left({{\cal R}_{\rm H}}^{-4}\right)
+ {\cal O}\left(N^{-2}\right) 
+ {\cal O}\left(N^{-1}{{\cal R}_{\rm H}}^{-2}\right) 
\ee
or 
\be
\label{RHsol}
{1 \over {\cal R}_{\rm H}^2}={4 \over 3N} + {\cal O}\left(N^{-2}\right) \ .
\ee
Thus, we demonstrated that next-to-leading order of AdS/CFT correspondence 
may qualitatively change the results on brane-world cosmology in
the leading order.
Indeed, in the leading order (Einstein theory) the warped compactification 
as 5d AdS with hyperbolic brane was impossible. On the same time,
account of next-to-leading terms (on the example of particular SCFT dual)
improves the situation: creation of hyperbolic brane in 5d AdS space 
becomes possible.

Let us discuss the situation where the higher derivative 
gravity in five  dimensions corresponds to the Gauss-Bonnet combination  
which is topological 
invariant in four deimensions. Then $a$, $b$ and $c$ are
given by
\be
\label{viGB}
a=c=\hat a \ , \quad b=-4\hat a\ .
\ee
One gets
\be
\label{GBT}
T_{\rm GB}=3\hat a\ .
\ee
Then all the discussion given above can be used  
by replacing $T$ by $3\hat a$ (compare with independent calculation in
ref.\cite{jekim}).  

The situation is changed for the case that the brane is R$_4$, 
where $A$ is given by
\be
\label{AR4}
A={z \over l} + \sigma\ .
\ee
Since $A_{,zz}=A_{,\sigma\sigma}=0$, the coefficient of 
$\delta A_{,z}$ in (\ref{breq}) is given by
\be
\label{RdAz}
0={32 T - 24 U - 96 \tilde T 
+ 72 \tilde U \over l^2}
- {8 \over \kappa^2} + {8 \over \tilde \kappa^2}
\ee
Then we obtain equations weaker than (\ref{wldeqs2}):
\be
\label{wldeqs2R}
\tilde U={4 \over 3}\tilde T\ ,\quad 
{1 \over \tilde\kappa^2}={1 \over \kappa^2} 
- {4T - 3U \over l^2}\ , 
\ee
and $\eta$ in (\ref{eta}) is given by
\be
\label{eta2H}
\eta={6 \over l\kappa^2} - {24T + 6U - 4\tilde T\over l^3}\ .
\ee
The brane equation, which is the coefficient of 
$\delta A$ in (\ref{breq}) has the following form:
\bea
\label{Rbreq}
0&=&\left(160 T - 128 \tilde T\right) {1 \over l^3} 
+\left(-{8 \over \kappa^2} + {32 \over \tilde\kappa^2}
\right){1 \over l} - 4\eta \nn
&=& {128T + 120U -  144\tilde T  \over l^3} \ .
\eea
Then we have
\be
\label{Rbreq2}
\tilde T={8 \over 9}T + {5 \over 6}U\ .
\ee
As one sees it admits the number of solutions for very large range of HD
terms
coefficients. Actually, chosing the suitable surface term the flat brane 
solution always exists.
Thus, we explicitly showed that brane-world cosmology with spherical or
hyperbolic
or flat brane is possible for big class of HD gravities.
The corresponding restrictions to HD terms coefficients are explicitly
obtained. The version of HD gravity corresponding to next-to-leading 
order of AdS/CFT correspondence for specific SCFT is naturally included as
sub-class of such theory.

\section{Brane-Worlds with Account of Brane Quantum Matter}

In the present section we will discuss the modification of the above
scenario in the situation when quantum matter lives on the brane.
Of course, bulk dynamics is not touched by brane quantum effects.
It is interesting to remark also that
in case of AdS/CFT correspondence the explanation of presence
of such quantum brane matter effective action naturally appears 
via holographic renormalization group \cite{ver}. In other words,
the two dual descritions (SG dual and QFT one) could be patched together 
into the unique global description of some RG flow \cite{ver}.
Of course, we will consider general situation when 
for general HD gravity with arbitrary coefficients 
some quantum CFT lives on the brane.

The quantum correction induced by the trace anomaly of 
the free conformally invariant matter fields on the brane can be realized
by adding the 
following effective action $W$ to $S+S_b$:
\bea
\label{W}
W&=& \hat b \int d^4x \sqrt{\widetilde g}\widetilde F A \nn
&& + b' \int d^4x \sqrt{\widetilde g}
\left\{A \left[2 {\widetilde{\Delta}}^2 
+\widetilde R_{\mu\nu}\widetilde\nabla_\mu\widetilde\nabla_\nu 
 - {4 \over 3}\widetilde R \widetilde\Delta^2 
+ {2 \over 3}(\widetilde\nabla^\mu \widetilde R)\widetilde\nabla_\mu
\right]A \right. \nn
&& \left. + \left(\widetilde G - {2 \over 3}\widetilde\Delta 
\widetilde R
\right)A \right\} \nn
&& -{1 \over 12}\left\{b''+ {2 \over 3}(b + b')\right\}
\int d^4x \sqrt{\widetilde g} \left[ \widetilde R 
- 6\widetilde\Delta A 
 - 6 (\widetilde\nabla_\mu A)(\widetilde \nabla^\mu A)
\right]^2 \ .
\eea
In (\ref{W}), one chooses the 4 dimensional boundary metric 
as 
\be
\label{tildeg}
\gfr_{\mu\nu}=\e^{2A}\tilde g_{\mu\nu}
\ee 
and we specify the 
quantities with $\tilde g_{\mu\nu}$ by using $\tilde{\ }$. 
$G$ ($\tilde G$) and $F$ ($\tilde F$) are the Gauss-Bonnet
invariant and the square of the Weyl tensor: 
\bea
\label{GF}
G&=&R^2 -4 R_{ij}R^{ij}
+ R_{ijkl}R^{ijkl} \nn
F&=&{1 \over 3}R^2 -2 R_{ij}R^{ij}
+ R_{ijkl}R^{ijkl} \ .
\eea
In the effective action (\ref{W}), with $N$ scalar, $N_{1/2}$ 
spinor, $N_1$ vector fields, $N_2$ ($=0$ or $1$) gravitons 
and $N_{\rm HD}$ higher 
derivative conformal scalars, $\hat b$, $b'$ and $b''$ are 
\bea
\label{bs}
\hat b&=&{N +6N_{1/2}+12N_1 + 611 N_2 - 8N_{\rm HD} 
\over 120(4\pi)^2}\nn 
b'&=&-{N+11N_{1/2}+62N_1 + 1411 N_2 -28 N_{\rm HD} 
\over 360(4\pi)^2}\ , 
\nn 
b''&=&0\ .
\eea
As usually, $b''$ may be changed by the finite 
renormalization of local counterterm in gravitational 
effective action. As we shall see later, the term proportional 
to $\left\{b''+ {2 \over 3}(\hat b + b')\right\}$ in (\ref{W}), and 
therefore $b''$, does not contribute to the equations of motion. 
For ${\cal N}=4$ $SU(N)$ SYM theory 
\be
\label{N4bb}
\hat b=-b'={N^2 -1 \over 4(4\pi )^2}\ ,
\ee 
and for ${\cal N}=2$ $Sp(N)$ theory 
\be
\label{N2bb}
\hat b={12 N^2 + 18 N -2 \over 24(4\pi)^2}\ ,\quad 
b'=-{12 N^2 + 12 N -1 \over 24(4\pi)^2}\ .
\ee
Notice that due to the structure of conformal anomaly the next-to-leading 
term for ${\cal N}=4$ super Yang-Mills theory dual is zero. 
Non-trivial term proportional to third power on curvatures 
appears in gravitational action as next-to-next-to-leading term.
We should also note that $W$ in (\ref{W}) is defined up to 
conformally invariant functional, which cannot be determined 
from only the conformal anomaly. The conformally flat space is 
an example where anomaly induced effective action 
is defined uniquely. However, one can argue that such 
conformally invariant functional is irrrelevant for us 
because it does not contribute to brane dynamics (does not depend on $A$).

In the choice of the metric (\ref{metric1}), we find 
$\tilde F=\tilde G=0$, $\tilde R={6 \over l^2}$ etc. and 
(\ref{W}) looks
\bea
\label{Wii}
W&=& V_3 \int d\sigma \left[b'A\left(2
A_{,\sigma\sigma\sigma\sigma}
 - 8 A_{,\sigma\sigma} \right) \right. \nn
&&\left. - 2(b + b')\left(1 - A_{,\sigma\sigma} 
 - (A_{,\sigma})^2 \right)^2 \right]\ .
\eea
Under the variation over $A$, the change of $W$ is given by 
\bea
\label{deltaW}
\delta W&=& V_3l^4 \int d\sigma \left\{4b'\left(
A_{,\sigma\sigma\sigma\sigma}-4 A_{,\sigma\sigma}\right)
- 4(\hat b+b')\left(A_{,\sigma\sigma\sigma\sigma}
+ 2 A_{,\sigma\sigma}
\right.\right.\nn
&& \left.\left. - 6 (A_{,\sigma})^2A_{,\sigma\sigma} \right)
\right\}\delta A \ .
\eea
Then by substituting the solution (\ref{AS4}), we find 
Eq.(\ref{breqc}) is changed as
\be
\label{breqc2}
0=2\left\{\left({24 \over \kappa^2} + {32 T\over l^2}
\right){\coth{z_0 \over l} \over l} - {24 \over l\kappa^2} 
+ {96 T\over l^3}\right\}\sinh^4{z_0 \over l} + 24 b'\ . 
\ee
Using the radius ${\cal R}$ of S$_4$, which is given in 
(\ref{tldR}), Eq.(\ref{breqc2}) is rewritten 
\be
\label{breqc3}
0=2\left\{\left({24 \over l\kappa^2} + {32 T\over l^3}
\right)\sqrt{1+ {l^2 \over {\cal R}^2}} - {24 \over l\kappa^2} 
+ {96 T\over l^3}\right\}{\cal R}^4 + 24b'\ . 
\ee
For  H$_4$ brane,  using the radius 
${\cal R}_{\rm H}$ in (\ref{tldRH}), one gets
\be
\label{breqc3H}
0=2\left\{\left({24 \over l\kappa^2} + {32 T\over l^3}
\right)\sqrt{1 - {l^2 \over {\cal R}_{\rm H}^2}} 
- {24 \over l\kappa^2} 
+ {96 T\over l^3}\right\} {\cal R}_{\rm H}^4 + 2b'\ . 
\ee
For R$_4$ brane, the equation (\ref{Rbreq}) or (\ref{Rbreq2}) 
is not changed. 

For the case of S$_4$, the l.h.s. of (\ref{breqc3}) goes to 
$24b'$ when ${\cal R}\rightarrow 0$ and behaves 
as ${256 T\over l^3}{\cal R}^4$ for large ${\cal R}$ if 
$T\neq 0$. Furthermore if $T>0$, the l.h.s. of (\ref{breqc3}) is 
the monotonically increasing function of ${\cal R}$. 
Then if $T>0$, $b'<0$, there is a unique solution. Since when $T>0$ 
there is no solution for the classical case in 
(\ref{breqc}), the solution for $T>0$ is generated by the 
quantum brane matter effects.  
We should note that even if $b'>0$ there is always a solution 
again if $T<0$. Here, HD gravity plays the essential role. On the other
hand even if $b'<0$ and $T<0$, 
the l.h.s. of (\ref{breqc3}) has a unique maximum as a 
function of ${\cal R}$. Then if the value of the maximum 
is positive, there are two solutions for ${\cal R}$, which 
satisfies (\ref{breqc3}). Since the l.h.s. of (\ref{breqc3}) is 
the monotonically increasing function of ${\cal R}$ when $T>0$, 
there is no any solution if $T>0$ and $b'>0$.

For the case of H$_4$, the situation does not  change if compare  
with  S$_4$ if $T\neq 0$. This is because the 
behavior of the r.h.s. in (\ref{breqc3H}) is again 
governed by the sign of $T$ when ${\cal R}_{\rm H}$ is large. 
Then if $b'<0$ and $T>0$ or if $b'>0$ and $T<0$, there is 
always a solution. The solution for $T<0$ is generated by the 
brane matter quantum effects. If $T<0$ and $b'<0$, there can be two 
(quantum) solutions. If $b'<0$ and $T<0$, there is no any solution. 

The interesting example is provided by 
 ${\cal N}=2$ $Sp(N)$ theory in the situation when SG dual and QFT 
descriptions are matched together via holographic RG, in both cases in
next-to-leading order 
of AdS/CFT correspondence. In other words,  
 using (\ref{N2ll}), (\ref{N2T}) (SG dual up to next-to-leading order)
and (\ref{N2bb}) (conformal anomaly for SCFT), in  Eqs. (\ref{breqc3}) and
(\ref{breqc3H}) leads to 
\bea
\label{breqCCCQ}
0&=&\left\{\sqrt{1+ {1 \over {\cal R}^2}} - 1
+ {1 \over 48 N}\sqrt{1+ {1 \over {\cal R}^2}}\left(5 
 - {1 \over {\cal R}^2 +1}\right) + {22 \over 48N}
\right\}{\cal R}^4 \nn
&& - {1 \over 384} - {1 \over 384 N} 
+ {\cal O}\left(N^{-2}\right) \ ,\\
\label{breqCCCHQ}
0&=&\left\{\sqrt{1 - {1 \over {{\cal R}_{\rm H}}^2}} - 1
+ {1 \over 48 N}\sqrt{1 - {1 \over {\cal R}^2}}\left(5 
- {1 \over {{\cal R}_{\rm H}}^2 - 1}\right) + {22 \over 48N}
\right\}{{\cal R}_{\rm H}}^4 \nn
&& - {1 \over 384} - {1 \over 384 N} 
+ {\cal O}\left(N^{-2}\right) \ . 
\eea
Since $T>0$ and $b'<0$, there are always solutions 
in (\ref{breqCCCQ}) and (\ref{breqCCCHQ}) for finite 
${\cal R}$ and ${\cal R}_{\rm H}$. 
When the brane is S$_4$ in (\ref{breqCCCQ}), ${\cal R}$ 
becomes ${\cal O}(1)$ and the higher derivative terms give 
a correction of ${\cal O}\left(N^{-1}\right)$. When 
$N\rightarrow \infty$, the solution of (\ref{breqCCCQ}) 
is numerically given by ${\cal R}^2=0.020833...$.  
Substituting this value into the r.h.s. of (\ref{breqCCCQ}), 
we find that it takes a negative value of $-{0.00215076 \over N}$. 
Since $T>0$, the r.h.s. is monotonically increasing function 
of ${\cal R}$ and goes negative when ${\cal R}\rightarrow 0$. 
Then the above result tells that the correction of 
${\cal O}\left(N^{-1}\right)$ makes ${\cal R}$ large. 
Since ${1 \over {\cal R}}$ corresponds to the rate of inflation 
when we Wick-rotate S$_4$ to de Sitter space, the correction 
makes the rate small. This is some indication that realistic inflationary 
cosmology may not be comfortable with warped compactification in AdS/CFT
correspondence.

For the brane of H$_4$, the quantum correction of 
${\cal O}(N^{-2})$ 
to ${1 \over {\cal R}_{\rm H}^2}$ of the classical solution 
in (\ref{RHsol}) exists  but since further higher derivative gravity 
terms like $R^4$ also give the contribution of 
${\cal O}(N^{-2})$, the correction is beyond the control.

Hence, we demonstrated that role of quantum brane matter may be
in the significant change of bulk/boundary structure.
As we saw there exists the range of HD terms coefficients for
which the creation of inflationary or hyperbolic Universe living 
in d5 AdS is caused exclusively by brane quantum effects. 
It could be also relevant in frames of AdS/CFT set-up where correct
holographic RG description shows the necessity of anomaly induced effective
action of brane CFT. In its own turn,  the corresponding quantum effects
change the brane structure and indicate (despite the negative results of  
leading
order analysis)
to the possibility of quantum creation of inflationary brane in d5 AdS
space
in the next-to-leading order of AdS/CFT correspondence.

\section{Discussion}

In summary, we investigated brane-world Universe solutions 
(of special form) for five-dimensional higher derivative gravity.
It is shown that such Universe occurs for range of theory parameters.
As brane part may be given by de Sitter space 
which after analytical continuation to Lorentzian signature represents 
ever expanding inflationary Universe 
then such configuration could be relevant to observable world.
The particular examples of Weyl, Gauss-Bonnet or SG dual to 
some SCFT are also examinated. The role of brane quantum CFT is
investigated
in the quantum creation of spherical or hyperbolic brane Universes.

There are few interesting topics which may be left for future studies.
First of all, one has to investigate the structure of HD propagator near 
brane in order to understand in detail how HD gravity is trapped. In other 
words, graviton profile and corrections to Newton potential should be
estimated. Second, the dilaton may be included into the analysis of this
paper. However, the number of HD terms in dilatonic gravity grows
significally.
As a result, the analysis is getting too complicated technically. 
Nevertheless, it could be done at least for some truncated versions of 
HD dilatonic gravity (say, conformally invariant theory or dilatonic 
Gauss-Bonnet). Note also that some versions of such theory may be
considered as SG duals for non-commutative (super) Yang-Mills theory
(presumbly in next-to-leading order). Third, other cosmologies may be
considered in the same fashion where bulk and (or) boundary is modified.
In particular, the situation where bulk is AdS black hole and boundary 
is some FRW Universe (or vice-versa) deserves careful study. 
Fourth, it would be interesting to discuss the cosmological perturbations 
around our background and the details of late-time inflation. For example, 
in Einstein gravity the domain wall CFT significally suppresses the metric 
perturbations \cite{HHR}. What will be the role of HD gravitational terms 
in such phenomenon?

\section*{Acknoweledgements} 
We thank O. Obregon and V. Tkach for helpful discussions.
The work by SDO has been supported 
in part by CONACyT (CP, ref.990356 and grant 28454E) 
and in part by RFBR.

\end{document}